\documentclass{article}
\usepackage{spconf,amsmath,amssymb,graphicx,bbm,mathtools,multirow,hhline,color,url,booktabs}

\def\jeit{{JEIT}}

\title{JEIT: Joint End-to-End Model and Internal Language Model Training for Speech Recognition}
\name{\begin{tabular}{c} Zhong Meng$^*$, Weiran Wang$^*$, Rohit Prabhavalkar, Tara N. Sainath, Tongzhou Chen, \\ Ehsan Variani, Yu Zhang, Bo Li, Andrew Rosenberg, Bhuvana Ramabhadran \end{tabular} \thanks{$^*$Equal Contribution.}}
\address{Google LLC., USA}
\begin{document}
\ninept
\maketitle
\begin{abstract}
We propose \jeit{}, a joint end-to-end (E2E) model and internal language model (ILM) training method to inject large-scale \emph{unpaired} text into ILM during E2E training which improves rare-word speech recognition. 
With \jeit{}, the E2E model computes an E2E loss on audio-transcript pairs while its ILM estimates a cross-entropy loss on unpaired text. The E2E model is trained to minimize a weighted sum of E2E and ILM losses. During \jeit{}, ILM absorbs knowledge from unpaired text while the E2E training serves as regularization.
Unlike ILM adaptation methods, \jeit{} does not require a separate adaptation step and 
avoids the need for Kullback-Leibler divergence regularization of ILM. 
We also show that modular hybrid autoregressive transducer (MHAT) performs better than HAT in the \jeit{} framework, and is much more robust than HAT during ILM adaptation.
To push the limit of unpaired text injection, we further propose a combined \jeit{} and JOIST training (CJJT) that benefits from modality matching, encoder text injection and ILM training. Both \jeit{} and CJJT can foster a more effective LM fusion. With 100B unpaired sentences, 
\jeit{}/CJJT improves rare-word recognition accuracy by up to 16.4\% over a model trained without unpaired text.
\end{abstract}
\section{Introduction}
\label{sec:intro}
End-to-end (E2E) models have achieved the strong performance for automatic speech recognition (ASR) \cite{chiu2018state, jain2019rnn, sainath2020streaming, li2020developing, zeyer2020new} by directly mapping the speech signal into word sequences. However, even trained with a large amount of audio-transcript pairs, the E2E models still performs poorly when evaluated on utterances including words that appear infrequently in the training data (rare words) \cite{sainath2021efficient}. Moreover, this supervised speech obtained via human transcription is expensive. To overcome this, utilizing knowledge from large-scale unpaired text during training or inference is a promising solution since unpaired text is orders of magnitude more plentiful than audio-transcript pairs and covers a much larger vocabulary of words. 

Language model (LM) fusion is a common approach to improve E2E ASR by using unpaired text. 
An external LM is first trained with unpaired text. In shallow fusion \cite{hannun2014deep,gulcehre2015on}, a log-linear interpolation between the E2E model score and the LM score is computed at each step of the beam search. 
To improve shallow fusion, internal LM estimation-based fusion \cite{mcdermott2019density, variani2020hybrid, meng2021ilme, zeyer2021librispeech, kanda2016maximum, meng2021minimum} was proposed to estimate an internal LM (ILM) score and subtract it from the shallow fusion score. 
However, all these methods require an external LM during inference, increasing decoding time and computational cost. 

To overcome this, various research has looked at incorporating unpaired text into the training stage of E2E models. One intuitive solution is to synthesize speech from unpaired text and use it to train the E2E model \cite{zhao2019shallow,rosenberg2019speech,deng2021improving,zheng2021using}.
However, training a text-to-speech (TTS) model and synthesizing speech are both computationally expensive. 
To circumvent this, modality matching approaches \cite{bapna2021slam,tang2022unified,thomas2022towards,chen2022maestro,sainath2022joist} were proposed to map unpaired text to a latent space shared by speech and text, and then use latent embeddings to train the E2E model.

Alternatively, the decoder (and joint network for a transducer model) of an E2E model behaves like an ILM when we zero out the encoder output \cite{variani2020hybrid, meng2021ilme, meng2021ilmt}. To achieve fast text-only adaptation, unpaired text is injected into the decoder of a well-trained E2E model \cite{pylkkonen2021fast, meng2021ilma, chen2022factorized, meng2022modular, gao2021pre}. 
These methods take one extra adaptation step of fine-tuning ILM of the E2E model
using text-only data to minimize a cross-entropy loss after one or two stages of E2E training.
In addition, Kullback-Leibler divergence (KLD) \cite{kullback1951information} regularization is performed to maintain the source-domain ASR performance.


The novel contributions of this work are: (1) We propose a joint E2E model and ILM training (\jeit{}) that simplifies decoder text injection by combining it into a single-stage E2E training. \jeit{} outperforms text-only adaptation without KLD regularization.
(2) We further propose a combined \jeit{} and JOIST \cite{sainath2022joist} training (CJJT) and demonstrate that decoder text-injection via ILM is complementary to encoder text-injection (via JOIST) and that the improvements are additive.
(3) We show that all text-injection methods can facilitate a more effective LM fusion.
(4) We validate our methods on Google's large-scale streaming production task where \jeit{} and CJJT offer up to 10.2\% and 16.4\% relative reductions in WER, respectively, on rare-word test sets without affecting voice search performance.


\section{Related Work}
\label{sec:format}

\subsection{Hybrid Autoregressive Transducer (HAT)}
\label{sec:hat}
An E2E model estimates the posterior distribution $P(\mathbf{Y} |
\mathbf{X};\theta_\text{E2E})$ over sequences of output labels $\mathbf{Y}=\{y_1, \ldots,
y_U\}$ given a sequence of input speech features $\mathbf{X}=\{\mathbf{x}_1,
\ldots, \mathbf{x}_T\}$, where $y_u \in \mathcal{V}, u = 1, \ldots, U$, and $\mathbf{x}_t
\in \mathbbm{R}^{d_x}, t = 1, \ldots, T$. $\mathcal{V}$ is the set of all possible labels, e.g., word pieces, etc. $y_0$ is the start of sentence token. 


HAT \cite{variani2020hybrid} consists of an acoustic encoder, a label decoder and a joint network. In Fig \ref{fig:jeit_hat}, the encoder transforms input speech features $\mathbf{X}$ into acoustic embedding vectors $\mathbf{F} = \{\mathbf{f}_1, \ldots, \mathbf{f}_T\}, \; \mathbf{f}_t \in \mathbbm{R}^{d_f}$, i.e., $\mathbf{F} = \text{Encoder}(\mathbf{X})$. The label decoder takes in previous labels to generate the current label embedding $\mathbf{g}^\text{L}_u \in \mathbbm{R}^{d^\text{L}_g}$, i.e., $\mathbf{g}^\text{L}_u = \text{LabelDecoder}(\mathbf{Y}_{0:u - 1})$. 
The joint network combines the acoustic and label embeddings to computes a blank distribution
\begin{align}
b_{t, u} = \text{Sigmoid}[\mathbf{w}^\intercal \phi(\mathbf{W}_1 \mathbf{f}_t + \mathbf{W}_2 \mathbf{g}^\text{L}_u)], \label{eqn:blank_posterior}
\end{align}
where $\mathbf{w} \in \mathbbm{R}^{d_h}$ is a vector, $\mathbf{W}_1 \in \mathbbm{R}^{d_h\times d_f}$ and $\mathbf{W}_2 \in \mathbbm{R}^{d_h\times d^\text{L}_g}$ are projection matrices.
$\phi(\cdot)$ is a non-linear function.
The label posteriors given previous speech features and labels are computed as
\begin{align}
P(y_u|\mathbf{X}_{1:t}, \mathbf{Y}_{0:u - 1}) = \text{Softmax}[\mathbf{W} \phi(\mathbf{W}_1 \mathbf{f}_t + \mathbf{W}_2 \mathbf{g}^\text{L}_u)], \label{eqn:label_posterior}
\end{align}
where $\mathbf{W} \in \mathbbm{R}^{|\mathcal{V}|\times d_h}$ is a projection matrix.
The label posterior given the alignment history is therefore $(1 - b_{t,u}) P(y_u|\mathbf{X}_{1:t}, \mathbf{Y}_{0:u - 1})$.

\subsection{Modular HAT (MHAT)}

To achieve more robust text-only adaptation, we proposed MHAT in \cite{meng2022modular}  to structurally separate the ILM score prediction from the acoustic model score or blank score predictions. 
As in Fig. \ref{fig:jeit_mhat}, MHAT introduces a blank decoder that takes in the same previous labels as the label decoder to generate the current label embeddings below 
\begin{align}
\mathbf{g}^\text{B}_u = \text{BlankDecoder}(\mathbf{Y}_{0:u - 1}),
\end{align}
where $\mathbf{g}^\text{B}_u \in \mathbbm{R}^{d^\text{B}_g}$ is the label embedding of the label decoder. 
The blank posterior $b_{t, u}$ is obtained by Eq. \eqref{eqn:blank_posterior} using $\mathbf{g}^\text{B}_u$.
$\mathbf{f}_t$ and $\mathbf{g}^\text{L}_u$ are projected and then normalized to be $|\mathcal{V}|$-dimensional (dim) vectors of AM log probabilities $\mathbf{a}_t$ and ILM log probabilities $\mathbf{l}_u$, respectively
\begin{align}
\mathbf{a}_t = \text{LogSoftmax}(\mathbf{W}_3 \mathbf{f}_t), \;\;
\mathbf{l}_u = \text{LogSoftmax}(\mathbf{W}_4 \mathbf{g}^\text{L}_u), \label{eqn:ilm_mhat}
\end{align}
where $\mathbf{W}_3 \in \mathbbm{R}^{|\mathcal{V}|\times d_f}$ and $\mathbf{W}_4 \in \mathbbm{R}^{|\mathcal{V}|\times d^\text{L}_g}$ are projection matrices.
$\mathbf{a}_t$ and $\mathbf{l}_u$ are added and then normalized to compute label posteriors
\begin{align}
P(y_u|\mathbf{X}_{1:t}, \mathbf{Y}_{0:u - 1}) = \text{Softmax}\left(\mathbf{a}_t + \mathbf{l}_u \right). \label{eqn:label_posterior_2}
\end{align}

\subsection{ILM Training (ILMT) and ILM adaptation (ILMA)}
ILMT \cite{variani2020hybrid, meng2021ilmt} minimizes an additional ILM loss during E2E model training. While the E2E loss is computed with audio-transcript pairs, the ILM loss is derived from only the training transcript. 
ILMT aims to encourage ILM to behave also like a standalone neural LM such that (1) accurate ILM scores can be estimated to improve ILME-based fusion \cite{meng2021ilmt} (2) ILM can be further adapted to text-only data \cite{meng2021ilma}. ILMT makes no use of unpaired text and it does not improve the ASR performance on either source-domain or rare-word test sets \cite{meng2021ilmt}. Unlike ILMT, \jeit{} injects \emph{unpaired} text into ILM during E2E training with the goal of improving rare-word recognition.



ILMA \cite{meng2021ilma} performs fast text-only adaptation of an E2E model to improve rare-word ASR.
In ILMA, we first conduct ILMT of E2E model and then fine-tune ILM to minimize a cross-entropy ILM loss using unpaired text. 
To prevent the source-domain ASR performance from degrading, we minimize an additional KLD between the output distributions of the unadapted and adapted ILMs during ILMA.
To simplify ILMA, \jeit{} combines two stages of ILMT and ILMA into one training stage and obviates the need for KLD regularization.

\subsection{Joint Speech and Text Modeling (JOIST)}
JOIST \cite{sainath2022joist} incorporates unpaired text into E2E training and significantly improves rare-word recognition. 
It injects unpaired text through the encoder so that text data can benefit the entire E2E model. 
In JOIST, unpaired text is first tokenized to word-piece or phoneme sequences and is then upsampled by replicating each token a fixed or random number of times. The upsampled text is masked and then fed into a text encoder to generate token embeddings which
are further passed to the decoder input or a layer of the encoder.
JOIST minimizes a weighted sum of two E2E losses derived from audio-transcript pairs $\mathcal{D}_\text{P}$ and unpaired text $\mathcal{D}_\text{UP}$, respectively
\begin{align}
    & \hspace{-0pt} \mathcal{L}_{\text{JOIST}}(\mathcal{D}_\text{P}, \mathcal{D}_\text{UP}) = \mathcal{L}_{\text{E2E}}(\mathcal{D}_\text{P}; \theta_\text{E2E}) + 
    \alpha \mathcal{L}_{\text{E2E}}(\mathcal{D}_\text{UP}; \theta_\text{E2E}), \label{eqn:ilmt}
\end{align}
where the two E2E losses are defined as
\begin{align}
    \hspace{-2pt}\mathcal{L}_{\text{E2E}}(\mathcal{D}_\text{P}; \theta_\text{E2E}) & = -\sum_{(\mathbf{X}, \mathbf{Y}) \in \mathcal{D}_\text{P}} \log P(\mathbf{Y}|\mathbf{X};\theta_\text{E2E}), \\
    \hspace{-0pt}\mathcal{L}_{\text{E2E}}(\mathcal{D}_\text{UP};\theta_\text{E2E}) & = - \hspace{-0pt}\sum_{\mathbf{Y} \in \mathcal{D}_\text{UP}} \log P\left(\mathbf{Y}|F(\mathbf{Y});\theta_\text{E2E}\right), \label{eqn:e2e_loss}
\end{align}
where $F(\cdot)$ is a function that tokenizes, unsamples and masks an unpaired sentence in $\mathcal{D}_\text{UP}$. $\alpha>0$ is the weight of unpaired E2E losses.
In this work, we incorporate ILM loss into JOIST to further improve ASR performance.

\section{Joint E2E and ILM Training (JEIT)}
ILM probability can be estimated by the E2E model output after zeroing out the encoder output \cite{variani2020hybrid, meng2021ilme}.
ILM is the decoder and the joint network of HAT, is the label decoder and output projection $\mathbf{W}_4$ of MHAT, and is the decoder of an AED model.

\begin{figure}[htpb!]
    \hspace{8pt}
	\includegraphics[width=0.8\columnwidth]{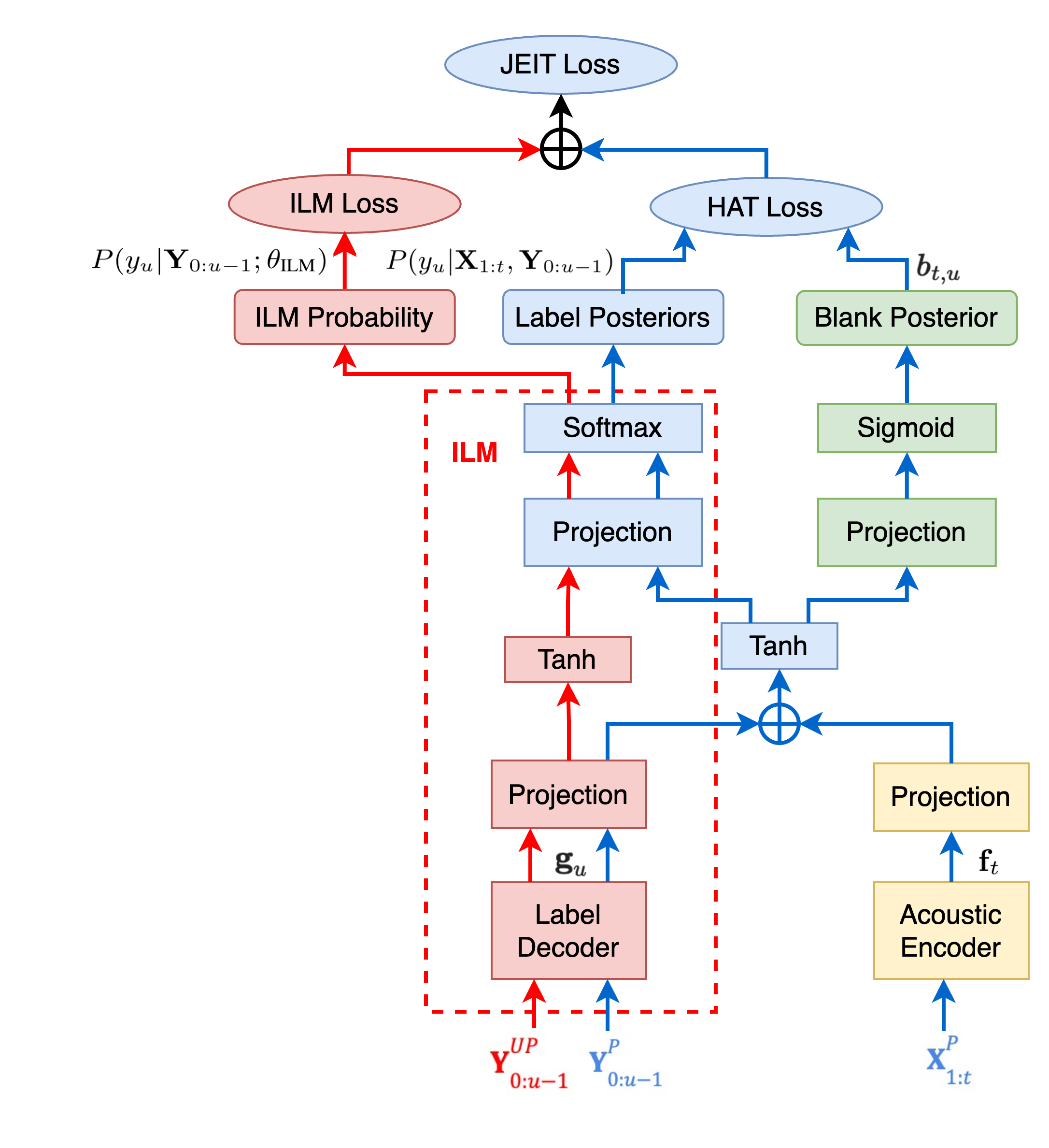}
    \vspace{-10pt}
    \caption{JEIT of HAT. Blue and red arrows represent the forward propagation path of audio-transcript pairs ($\mathbf{X}^\text{P}_{1:t}$, $\mathbf{Y}^\text{P}_{0:u-1}$) and unpaired text $\mathbf{Y}^\text{UP}_{0:u-1}$, respectively.} 
	\label{fig:jeit_hat}
\end{figure}
Our goal is to improve the ASR accuracy on rare-word test sets by making use of large-scale \emph{unpaired text} while maintaining WER on source-domain task (e.g., voice search). 
In this work, we propose \jeit{}, a joint training of E2E model and ILM that injects unpaired text into ILM during E2E training. 
\begin{figure}[htpb!]
    \hspace{0pt}
	\includegraphics[width=0.9\columnwidth]{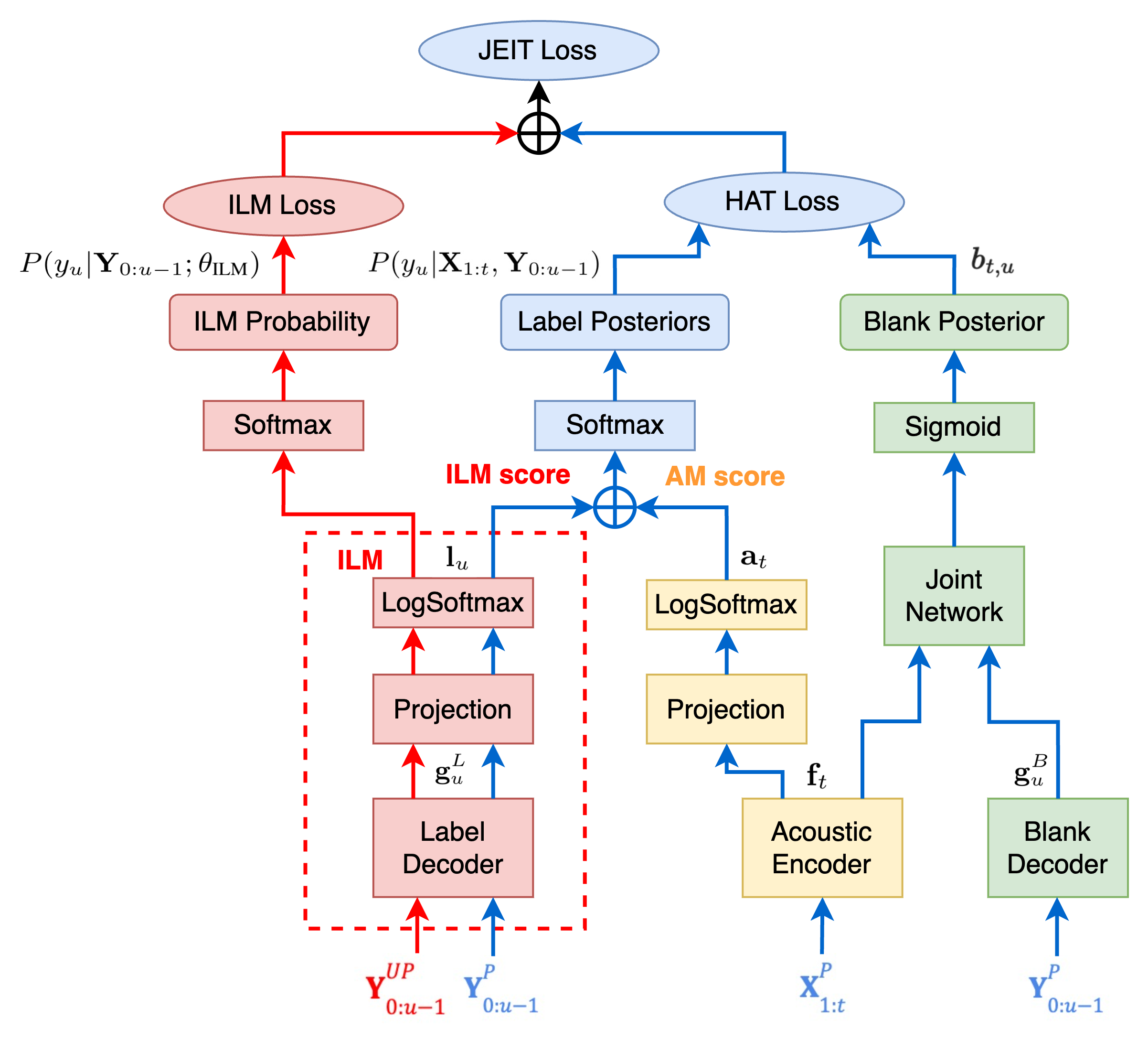}
    \vspace{-10pt}
    \caption{\jeit{} of MHAT. Blue and red arrows represent the forward propagation path of audio-transcript pairs ($\mathbf{X}^\text{P}_{1:t}$, $\mathbf{Y}^\text{P}_{0:u-1}$) and unpaired text $\mathbf{Y}^\text{UP}_{0:u-1}$, respectively.}
	\label{fig:jeit_mhat}
\end{figure}
As shown in Figs. \ref{fig:jeit_hat} and \ref{fig:jeit_mhat}, ILM is trained with \emph{unpaired} text to minimize an ILM loss while the entire E2E model is trained with audio-transcript pairs to minimize an E2E loss. The ILM loss minimization makes ILM a strong neural LM in the target domain while the E2E loss serves as a regularization to ensure ILM can work well with the other E2E model components to predict accurate E2E scores.
Specifically, \jeit{} minimizes a weighted sum of the E2E loss and ILM loss below
\begin{align}
    & \hspace{-0pt} \mathcal{L}_{\text{\jeit{}}}(\mathcal{D}_\text{P}, \mathcal{D}_\text{UP}) = \mathcal{L}_{\text{E2E}}(\mathcal{D}_\text{P};\theta_\text{E2E}) + 
    \beta \mathcal{L}_{\text{ILM}}(\mathcal{D}_\text{UP};\theta_\text{ILM}), \label{eqn:ilmt}
\end{align}
where $\beta > 0$ is the weight of the ILM loss. The ILM loss is the summed negative log probability of all label sequences predicted by ILM on the unpaired text $\mathcal{D}_\text{UP}$ as follows
\begin{align}
    \hspace{-2pt}\mathcal{L}_{\text{ILM}}(\mathcal{D}_\text{UP};\theta_\text{ILM}) 
    =-\sum_{\mathbf{Y} \in \mathcal{D}_\text{UP}} \sum^{U}_{u=1}\log P(y_u|\mathbf{Y}_{0:u-1};\theta_\text{ILM}),
    \label{eqn:ilm_loss_train}
\end{align}
where $\theta_\text{ILM} \subseteq \theta_\text{E2E}$ denotes ILM parameters.

Compared to text-only adaptation \cite{pylkkonen2021fast, meng2021ilma, chen2022factorized, meng2022modular}, \jeit{} significantly simplifies the entire learning process: 1) \jeit{} reduces two steps of audio-transcript training and unpaired text adaptation to one step of joint training, decreasing the computational cost and training/adaptation time. 2) \jeit{} avoids the need for KLD regularization of the ILM output distribution.

To improve rare-word recognition, \jeit{} injects unpaired text into the label decoder of an E2E model to minimize $\mathcal{L}_{\text{E2E}}(\mathcal{D}_\text{P};\theta_\text{E2E})$ and $\mathcal{L}_{\text{ILM}}(\mathcal{D}_\text{UP};\theta_\text{ILM})$ while JOIST injects it through the encoder to minimize $\mathcal{L}_{\text{E2E}}(\mathcal{D}_\text{P};\theta_\text{E2E})$ and $\mathcal{L}_{\text{E2E}}(\mathcal{D}_\text{UP};\theta_\text{E2E})$. To benefit from both methods, we proposed a combined \jeit{} and JOIST training (CJJT) to minimize a weighted sum of an E2E loss derived from audio-transcript pairs, an E2E loss derived from unpaired text and an ILM loss derived from unpaired text as follows
\begin{align}
    \hspace{-0pt} \mathcal{L}_{\text{CJJT}}(\mathcal{D}_\text{P}, \mathcal{D}_\text{UP}) &= \mathcal{L}_{\text{E2E}}(\mathcal{D}_\text{P};\theta_\text{E2E}) + \alpha \mathcal{L}_{\text{E2E}}(\mathcal{D}_\text{UP};\theta_\text{E2E}) \nonumber \\ 
    & + \beta \mathcal{L}_{\text{ILM}}(\mathcal{D}_\text{UP};\theta_\text{ILM}). \label{eqn:ilmt}
\end{align}
We show in the experiments that \jeit{} and JOIST are complementary to each other and CJJT achieves better ASR performance than either method alone.

During inference, we can integrate an external LM into the E2E model after \jeit{} or CJJT to further improve the rare-word ASR. We show that LM fusion is complementary to both \jeit{} and CJJT even if the external LM is trained with same unpaired text $\mathcal{D}_\text{UP}$ as in \jeit{}. 


\section{Experiments}

\subsection{Dataset}
\label{sec:production_data}
We use $\sim$650M multi-domain English audio-transcript pairs as supervised training data \cite{sainath2022joist}. It covers multiple domains including Voice Search, Dictation, YouTube, Telephony and etc. YouTube transcripts are generated in a semi-supervised fashion \cite{liao2013large} while other data is anonymized and hand-transcribed \cite{googleai}.
In addition, multi-condition training \cite{kim2017generation}, random 8kHz down-sampling \cite{li2012improving} and SpecAug \cite{park2019specaugment} are applied to augment and diversify the data. 

The unpaired text used in training or adaptation consists of 100B anonymized sentences across the domains of Maps, Google Play, Web, and YouTube, and is more than two orders of magnitude larger than audio-transcript pairs. The external LM is trained with 50\% transcripts of the paired data and 50\% unpaired text to ensure the quality on base Voice Search task does not degrade.

We evaluate our models on a Voice Search (VS) test set containing $\sim$12K anonymized and hand-transcribed voice search utterances with an average duration of 5.5 s. To evaluate ASR performance on long-tail words, we construct rare-word test sets for each of the 5 domains: Maps, Google Play, Web and YouTube (YT) domains. All test sets include rare proper nouns that appear fewer than 5 times in the training set and are synthesized by a TTS system \cite{gonzalvo2016recent}.
Our goal is to improve the ASR accuracy on 4 rare-word test sets without degrading the WER on Voice Search.
\vspace{-1pt}

\subsection{Modeling}
\label{sec:model}
We train HAT and MHAT with 2-pass cascaded encoders and separate decoders as in \cite{ding2022unified, narayanan2021cascaded}. They share the same front-end and encoder architecture.
Specifically, 128-dim log Mel filterbanks are extracted from speech signal and are subsampled to form a 512-dim feature every 30 ms.
Each speech feature is appended with a 16-dim domain ID \cite{narayanan2019recognizing}. 
The causal encoder is a 7-layer conformer with causal convolution and left-context attention. 
The non-causal encoder is a 10-layer conformer with right-context attention that processes 900 ms of speech into the future. Each conformer layer uses a 512-dim 8-head self-attention and a convolution kernel of size 15.

The causal and non-causal decoders of HAT or MHAT decode using the outputs of the causal and non-causal encoders, respectively. The label decoders of HAT and MHAT are 2-layer LSTMs with 2048 hidden units in each layer. 
In HAT, the label decoder output passes through a 640-dim feedforward joint network before projected to 4096 output units representing word pieces \cite{schuster2012japanese}. In MHAT, the label decoder output is directly projected to the output layer of the same size. 
ILMs of HAT and MHAT have 30.7M and 30M parameters, respectively. The blank decoder of MHAT is a 320-dim $V^2$ embedding decoder \cite{botros2021tied, ghodsi2020rnn} with a look-up table shared between the last 2 tokens and has 1.5M parameters. 
Overall, HAT and MHAT have in total 205M and 210M model parameters, respectively. We report only the 2nd pass WER in this paper. We train baselines with only audio-transcript pairs and show their WERs in Table \ref{table:ilma_jeit}.

Moreover, we train a 12-layer conformer LM with 384-dim self-attention and 3072-dim feedforward layer \cite{sainath2021efficient}. The external LM has left attention context of 31 and has in total 70M parameters. 

\subsection{ILMA of HAT and MHAT}
\begin{figure}[htb]
\vspace{-2pt}
\begin{minipage}[b]{0.48\linewidth}
  \centering
  \centerline{\includegraphics[width=4.48cm]{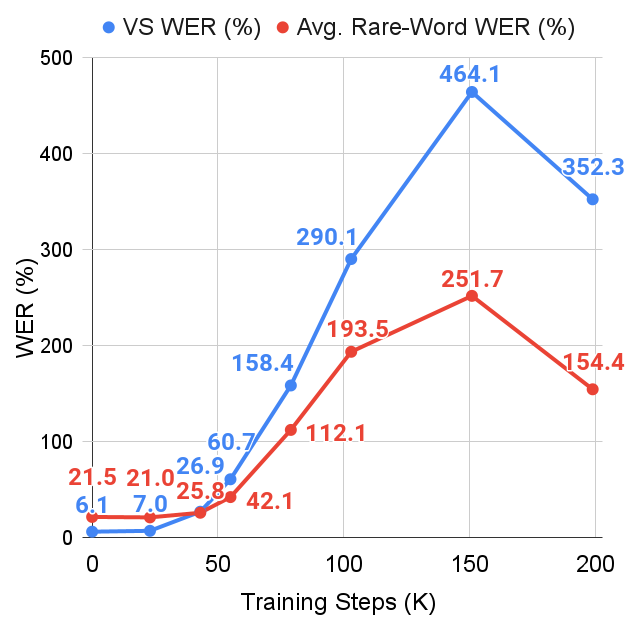}}
  \label{fig:ilma_hat}
  \centerline{(a) HAT ILMA}\medskip
\end{minipage}
\hfill
\begin{minipage}[b]{0.48\linewidth}
  \centering
  \centerline{\includegraphics[width=4.48cm]{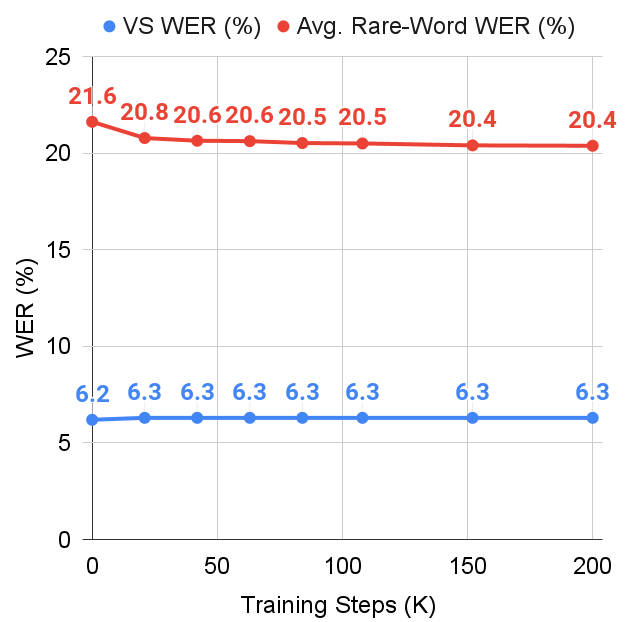}}
  \centerline{(b) MHAT ILMA}\medskip
  \label{fig:ilma_mhat}
\end{minipage}
\vspace{-13pt}
\caption{WERs (\%) of ILMA at different number of training steps (K) for HAT and MHAT with LSTM label decoders.}
\label{fig:ilma}
\end{figure}

We first train an ILMT \cite{meng2021ilmt} model with an ILM loss weight of 0.1 and use it as the seed for ILMA \cite{meng2021ilma}.
For both ILMA and \jeit{}, we adopt minibatch sizes of 4,096 and 32,768 for paired audio-transcript data and unpaired text, respectively. During ILMA, a KLD regularization with a weight of $0.5$ is applied for both HAT and MHAT. 
In Fig. \ref{fig:ilma}, we plot the WERs of HAT ILMA and MHAT ILMA with respect to number of training steps. WER of HAT ILMA sharply increases after reaching its best one at 5K training step while the WER of MHAT gradually decreases until after 200K step.  This is because without a structural factorization, HAT is not able to work with an increasingly stronger ILM and will lose its functionality of performing E2E ASR. This shows that MHAT is superior to HAT for ILMA because its structurally independent ILM
allows MHAT to constantly improve its ASR capability as ILM becomes stronger. We list the best WERs of ILMA in Table \ref{table:ilma_jeit}.

\vspace{-5pt}

\subsection{\jeit{} of HAT and MHAT}
We perform \jeit{} with the same minibatch sizes as ILMA and using ILM loss weights $\beta$ of 0.2 and 4.0 for HAT and MHAT, respectively. Significantly larger optimal ILM loss weight for MHAT signifies its advantage over HAT due to factorization - MHAT can work with increasingly stronger ILM to perform better ASR while HAT cannot.  
In Table \ref{table:ilma_jeit}, MHAT \jeit{} performs the best among all methods, achieving 4.8\%--10.2\% relative WER reduction from the baseline HAT on rare-word test sets.
For MHAT, \jeit{} gets better WERs than ILMA on all test sets.
MHAT \jeit{} consistently outperforms HAT \jeit{} by up to 3.5\% relatively in terms of lower WER. 
As \jeit{} goes on, WERs of both HAT and MHAT reduce continuously without any sudden increase.
This implies that E2E loss in \jeit{} serves as a much better regularization than KLD in ILMA.
Overall, we show for the first time that joint training of ILM is better than adaptation.

\begin{table}[h]
\centering
\setlength{\tabcolsep}{3.8pt}
\begin{tabular}[c]{c|c|c|c|c|c|c}
	\hline
	\hline
	\multirow{2}{*}{\begin{tabular}{@{}c@{}} Model \end{tabular}} &
	\multirow{2}{*}{\begin{tabular}{@{}c@{}} Exp \end{tabular}} & \multirow{2}{*}{\begin{tabular}{@{}c@{}} VS \hspace{0pt} \end{tabular}} & \multicolumn{4}{c}{Rare Words} \\
	\hhline{~~~----}
	& & & \hspace{-1pt} Maps \hspace{-1pt} & \hspace{1pt} Play \hspace{0pt} & \hspace{0pt} Web \hspace{0pt} & \hspace{1pt} YT \hspace{1pt} \\
	\hline
	\multirow{4}{*}{\begin{tabular}{@{}c@{}} HAT \end{tabular}} &
	Base & \textbf{6.1} & 14.0 & 37.3 & 21.6 & 24.6 \\
	& ILMT & 6.1 & 14.1 & 37.6 & 21.7 & 24.8 \\
	& ILMA & 6.2 & 13.6 & 36.3 & 20.8 & 24.2 \\
	& \jeit{} & 6.3 & 13.3 & 36.8 & 20.1 & 23.2 \\
	\hline
	\multirow{4}{*}{\begin{tabular}{@{}c@{}} MHAT \end{tabular}} &
    Base & 6.2 & 14.1 & 37.4 & 21.4 & 24.6 \\
	& ILMT & 6.2 & 14.3 & 37.8 & 21.8 & 25.0 \\
	& ILMA & 6.3 & 13.3 & 35.5 & 20.3 & 23.5 \\
	& \jeit{} & 6.2	& \textbf{13.2} & \textbf{35.5} & \textbf{19.4} & \textbf{22.6} \\
	\hline
	\hline
	\end{tabular}
	\caption{WERs (\%) of HAT, MHAT with LSTM label decoders using various training or adaptation methods.
	Baseline and ILMT \cite{meng2021ilmt} models are trained with 650M multi-domain (MD) audio-transcript pairs. The same paired data and 100B MD unpaired sentences are used for ILMA \cite{meng2021ilma} and the proposed \jeit{}.} 
\label{table:ilma_jeit}
\vspace{-10 pt}
\end{table}

\subsection{\jeit{} with Different Decoders}
We vary the type and size of the MHAT label decoders while keeping cascaded encoders in Section \ref{sec:model} unchanged. 
Besides LSTM, we explore simpler and smaller label decoders: $V^2$ embedding and $V^4$ embedding \cite{botros2021tied} which have 640-dim embeddings and condition on the last 2 and 4 tokens, respectively. Each previous token has a separate look-up table. The same blank decoder in Section \ref{sec:model} is used.
MHATs with $V^2$ and $V^4$ embedding decoders have 8.6M, 14.6M parameters for their ILMs and have in total 169M, 182M parameters, respectively.
In Tables \ref{table:jeit_decoder} and \ref{table:ilma_jeit}, \jeit{} of MHATs with $V^2$ embedding, $V^4$ embedding and LSTM decoders achieve 1.6\%--4.1\%, 1.6\%-4.9\% and  4.3\%--8.8\% relative WER reductions from the baseline MHAT on Maps, Play, Web and YT, respectively. 
For all 3 decoders, \jeit{} obtains no WER degradation on rare-word test sets. This shows that \jeit{} is beneficial to label decoders of various types and sizes. The effectiveness of \jeit{} increases as the ILM size grows and also as the label decoder's conditioning history extends.


\begin{table}[h]
\centering
\setlength{\tabcolsep}{3.0pt}
\begin{tabular}[c]{c|c|c|c|c|c|c|c}
	\hline
	\hline
	\multirow{2}{*}{\begin{tabular}{@{}c@{}} Label \\ Dec \end{tabular}} &
	\multirow{2}{*}{\begin{tabular}{@{}c@{}} Params \\ (M) \end{tabular}} &
	\multirow{2}{*}{\begin{tabular}{@{}c@{}} Exp \end{tabular}} & \multirow{2}{*}{\begin{tabular}{@{}c@{}} VS \hspace{0pt} \end{tabular}} & \multicolumn{4}{c}{Rare Words} \\
	\hhline{~~~~----}
	& & & & \hspace{-1pt} Maps \hspace{-1pt} & \hspace{1pt} Play \hspace{0pt} & \hspace{0pt} Web \hspace{0pt} & \hspace{1pt} YT \hspace{1pt} \\
	\hline
	\multirow{2}{*}{\begin{tabular}{@{}c@{}} $V^2$ \\ Embed \end{tabular}} &
	\multirow{2}{*}{\begin{tabular}{@{}c@{}} 8.6 \end{tabular}} &
	Base & 6.2 & 14.5 & 37.9 & 21.9 & 24.9 \\
	& & \jeit{} & 6.3 & 13.9 & \textbf{36.6} & 21.3 & 24.5 \\
	\hline
	\multirow{2}{*}{\begin{tabular}{@{}c@{}} $V^4$ \\ Embed \end{tabular}} &
	\multirow{2}{*}{\begin{tabular}{@{}c@{}} 14.6 \end{tabular}} &
	Base & 6.3 & 14.4 & 37.5 & 22.1 & 25.0 \\
	& & \jeit{} & \textbf{6.2} & \textbf{13.7} & 36.9 & \textbf{21.1} & \textbf{24.3} \\
	\hline
	\hline
	\end{tabular}
	\caption{WERs (\%) of MHATs with $V^2$ embedding, $V^4$ embedding decoders.
	Cascaded encoders and blank decoders of the two MHATs are the same as those of MHAT with LSTM label decoder in Table \ref{table:ilma_jeit}}
\label{table:jeit_decoder}
\end{table}
\vspace{-10 pt}

\subsection{Combining \jeit{} with Other Text Injection Methods}
We train JOIST MHAT with phoneme-based unpaired text following the setup in \cite{sainath2022joist}. 
The text encoder output is fed to the 3rd conformer layer of causal encoder. Unpaired E2E loss weight $\alpha$ is 0.25. 
We conduct combined \jeit{} and JOIST training (CJJT) with an ILM loss weight $\beta$ of 1.5. We subtract ILM scores during LM fusion. 


\begin{table}[h]
\centering
\setlength{\tabcolsep}{2.5pt}
\begin{tabular}[c]{c|c|c|c|c|c|c}
	\hline
	\hline
	\multirow{2}{*}{\begin{tabular}{@{}c@{}} Exp \end{tabular}} &
	\multirow{2}{*}{\begin{tabular}{@{}c@{}} Params \\ (M) \end{tabular}} &
	\multirow{2}{*}{\begin{tabular}{@{}c@{}} VS \hspace{0pt} \end{tabular}} & \multicolumn{4}{c}{Rare Words} \\
	\hhline{~~~----}
	& & & \hspace{-1pt} Maps \hspace{-1pt} & \hspace{1pt} Play \hspace{0pt} & \hspace{0pt} Web \hspace{0pt} & \hspace{1pt} YT \hspace{1pt} \\
	\hline
    Base & 
	\multirow{4}{*}{\begin{tabular}{@{}c@{}} 210 \end{tabular}} &
	6.2 & 14.1 & 37.4 & 21.4 & 24.6 \\
	\jeit{} &
	& 6.2 & 13.2 & 35.5 & 19.4 & 22.6 \\
	\hhline{-~-----}
	JOIST & 
	& 6.2 & 12.4 & 34.7 & 18.5 & 22.1 \\
	\jeit{} + JOIST (CJJT)  & 
	& 6.2 & 12.0 & 33.8 & 17.9 & 21.3 \\
	\hline
	Base + LM & 
	\multirow{4}{*}{\begin{tabular}{@{}c@{}} 280 \end{tabular}}
    & 6.0 & 11.8 & 34.4 & 17.9 & 23.2 \\
	\jeit{} + LM & 
    & 6.0 & 11.7 & 34.0 & 17.0 & 22.6 \\
	\hhline{-~-----}
	\multirow{2}{*}{\begin{tabular}{@{}c@{}} CJJT + LM \\ CJJT + MWER + LM \end{tabular}} &
& \textbf{6.0} & 10.6 & 31.9 & 15.6 & 20.8 \\
	\hhline{~~~~~~~}
	& & 6.1 & \textbf{10.1} & \textbf{30.7} & \textbf{14.7} & \textbf{19.3} \\
	\hline
	\hline
	\end{tabular}
	\caption{WERs (\%) of MHATs with LSTM label decoders when combining \jeit{} with various text injection methods and MWER.}
\label{table:jeit_combine}
\vspace{-5 pt}
\end{table}

CJJT consistently outperforms both JOIST and \jeit{}, indicating text injection into the encoder and decoder are complementary and the gains are additive. 
It is worth noting that CJJT achieves similar or even better WER than LM fusion with base MHAT on rare-word test sets, despite having 70M fewer model parameters.
LM fusion with \jeit{}/CJJT MHAT achieves 2.3\%--12.8\% \emph{additional} gains relatively, 
so conducting \jeit{} or CJJT early on is extremely beneficial to LM fusion. LM fusion with CJJT performs better than with \jeit{}, suggesting \jeit{}, JOIST and LM fusion are complementary to each other.
Finally, we perform CJJT, minimum word error rate (MWER) training \cite{prabhavalkar2018minimum} and LM fusion, and obtain the best WER over all systems with 17.9\%--31.3\% relative WER reductions from the baseline.

\section{Conclusion}
We propose \jeit{} to inject unpaired text into ILM via a single-stage joint training.
\jeit{} simplifies two-stage ILMA and eliminates KLD regularization, achieving up to 10.2\% relative WER reductions from baseline on rare-word test sets. 
MHAT performs better than HAT after \jeit{}, and is much more robust than HAT during ILMA. Text injection into encoder and decoder are complementary, combining them (CJJT) achieves up to 16.4\% relative gain. LM fusion further improves all text-injection methods by up to 12.8\% relatively. 
\vfill\pagebreak

\bibliographystyle{IEEEbib}
\bibliography{refs}

\end{document}